\documentclass[prd,aps,twocolumn,a4paper,nofootinbib]{revtex4-1}

\newif\ifusesec
\usesectrue 
   
\usepackage{graphicx,psfrag}
\usepackage{mathrsfs}
\usepackage{amsmath,amsfonts,amssymb}
\usepackage{multirow}
\usepackage{comment}

\newcommand{\beq}{\begin{equation}}
\newcommand{\eeq}{\end{equation}}

\begin{document}

\title{Analytic determination of the eight-and-a-half post-Newtonian   self-force contributions to the two-body gravitational interaction potential}

\author{Donato \surname{Bini}$^1$}
\author{Thibault \surname{Damour}$^2$}

\affiliation{$^1$Istituto per le Applicazioni del Calcolo ``M. Picone'', CNR, I-00185 Rome, Italy\\
$^2$Institut des Hautes Etudes Scientifiques, 91440 Bures-sur-Yvette, France}

\date{\today}

\begin{abstract}
We {\it analytically} compute, to the eight-and-a-half post-Newtonian order, and to linear order in the mass ratio, the 
radial potential describing (within the effective one-body formalism) the gravitational interaction of two bodies, thereby
extending  previous analytic results.
These results are obtained by applying analytical gravitational self-force theory (for a particle in circular orbit
around a Schwarzschild black hole) to Detweiler's gauge-invariant redshift variable.
We emphasize the increase in \lq\lq transcendentality" of the numbers entering the post-Newtonian expansion coefficients as the order increases, in particular
we note the appearance of $\zeta(3)$ (as well as the square of Euler's constant $\gamma$) starting at the seventh post-Newtonian order.
We study the convergence of the  post-Newtonian  expansion as the expansion parameter $u=GM/(c^2r)$ leaves the weak-field domain $u\ll 1$ to enter the strong field domain $u=O(1)$.
\end{abstract}


\maketitle

\section{Introduction}
\label{sec:intro}

This paper is the third in a sequence of works \cite{Bini:2013zaa,bdtd_beyond} devoted to the analytic  determination of the main radial potential $A(r; m_1,m_2)$  describing  (in a gauge-invariant way) the gravitational interaction of two bodies, of masses $m_1,m_2$, within the effective one-body (EOB) formalism  \cite{Buonanno:1998gg,Buonanno:2000ef,Damour:2000we,Damour:2001tu}. [The function $A(r; m_1,m_2)$ represents (minus) the $g_{00}$ component of the effective metric entering the EOB formalism, thereby generalizing the well-known Schwarzschild function $A(r; m_1=0,m_2)=1-2Gm_2/(c^2r)$.]

In Ref. \cite{Bini:2013zaa} we completed the analytic determination of the radial potential $A(r; m_1,m_2)$ at the fourth post-Newtonian (4PN) approximation, without making any smallness assumption about the symmetric mass ratio $\nu = m_1 m_2 /(m_1+m_2)^2$. In Ref. \cite{bdtd_beyond}, we analytically determined up to the sixth post-Newtonian (6PN) order the contributions to $A(r;m_1,m_2)=A(u; \nu)$ [where $u:=G(m_1+m_2)/(c^2r)$] that are {\it linear} in $\nu$. Here we extend the analytic results of 
\cite{Bini:2013zaa} to the eight-and-a-half post-Newtonian (8.5PN) order, still working linearly in the symmetric mass ratio $\nu$. In the following, we shall not repeat the details of our framework (which have been expounded in \cite{bdtd_beyond}), but only recall the few technical facts that we need to present our new results.

We consider a two-body system with masses $m_1$ and $m_2$ in motion along a circular
orbit of (areal) radius $r_0$ and orbital frequency $\Omega$, in the
limit where $m_1 \ll m_2$. 
We denote by: $M=m_1+m_2$  the total mass of the system, $\mu=m_1m_2/(m_1+m_2)$ its reduced mass and 
$\nu =\mu/M= m_1 m_2 /(m_1+m_2)^2$ its symmetric mass ratio. 
In the present paper, we work (like in \cite{bdtd_beyond}) to {\it linear order} in $\nu$. To be definite, we consider that $m_1\ll m_2$.

The small mass $m_1$ perturbs the Schwarzschild spacetime metric associated with the mass $m_2$, $g_{\mu\nu}^{\rm Schw} (x;m_2)$
so that  $\delta g_{\mu\nu}= g_{\mu\nu} (x^\lambda ;m_1,m_2) - g_{\mu\nu}^{\rm Schw} (x^\lambda ;m_2) =(m_1/m_2) h_{\mu\nu}(x^\lambda) + O(m_1^2/m_2^2)$. 
Techniques have been developed over the years to analytically compute the (rescaled) metric perturbation $h_{\mu\nu}(x^\lambda)$ (see references in \cite{bdtd_beyond}). As emphasized by Detweiler \cite{Detweiler:2008ft}, an interesting {\it gauge-invariant} quantity associated with $h_{\mu\nu}$ is the function
\beq
\label{h_kk_reg}
 h_{kk}^{R}(u) := h_{\mu\nu}^{R} k^{\mu} k^{\nu} \,.
 \eeq
Here $k^{\mu}$ denotes  the helical Killing vector $k^{\mu} \partial_{\mu} = \partial_t + \Omega \partial_{\varphi}$,   and the superscript $R$
 denotes the   {\it regularized}  value of  $h_{\mu\nu}(x^\lambda)$ on the world line of the small mass $m_1$.  In addition, 
 the argument $u$ in the function  $ h_{kk}^{R}(u) $ denotes  $G M/c^2 r_0$ or, equivalently, 
 $(GM\Omega/c^3)^{2/3}$. [The argument of the first-order metric perturbation $(m_1/m_2)h_{kk}(u)$ needs only to be defined with background accuracy.] 

The explicit analytic computation of $ h_{kk}^{R}(u)$ requires both the improved analytic black hole perturbation techniques developed by the Japanese relativity school \cite{Mano:1996vt,Mano:1996mf,Mano:1996gn} and the spherical-harmonics-mode-sum regularization (see e.g. Refs. \cite{Barack:1999wf,Hikida:2004hs,Hikida:2004jw,Detweiler:2008ft}). In Ref. \cite{Bini:2013zaa}, it was enough to use the hypergeometric-expansion technique of Mano, Suzuki and Tagasugi for the quadrupolar ($l=2$) contribution to $h_{kk}(u)$. In Ref. \cite{bdtd_beyond}, we used this technique for the multipole orders $l=2,3$ and $4$. Indeed, this technique is needed to capture the infrared-delicate contributions induced by tail-related hereditary near-zone effects, and it was shown in \cite{bdtd_beyond} that the hereditary contribution from the $l$-th multipole starts at the $(l+2)$-th PN order. As our aim here is to reach the $8.5$PN order (neglecting 9PN and higher) we used the hypergeometric-expansion technique (followed by an expansion in powers of $1/c$ up to $1/c^{17}$ included) for $l=2,3,4,5$ and $6$. This necessitated to solve the three-term recursion equation determining the hypergeometric coefficients $a_n^\nu$ for $-20<n<+20$ (i.e., setting $a_{-20}^\nu=0=a_{20}^\nu$). In addition, we had to extend the determination of the PN-expanded solutions used for $l>6$ to the order $1/c^{17}$.

There are several ways to present our results. A first way, would be to exhibit the PN expansion of the quantity we actually computed, namely $h_{kk}^R(u)$, Eq. (\ref{h_kk_reg}). However, it was shown in Refs. ~\cite{Tiec:2011ab,Tiec:2011dp,Barausse:2011dq} that the function $h_{kk}^R(u)$ is very simply related to
the   (gauge-invariant) EOB 
 radial interaction potential $A(r;m_1,m_2)= A(u;\nu)$ of the two bodies via
  \beq
  \label{avshkk}
 h_{kk}^{R}(u)= -2a_{1\rm{SF}}(u)+ \xi(u)  \, ,
 \eeq
where
\begin{eqnarray}
\label{xi_exp}
\xi(u)&=& -2\frac{u(1-4u)}{\sqrt{1-3u}}\nonumber\\
&=&  -2 u+5 u^2+\frac{21}{4} u^3+\frac{81}{8} u^4+\frac{1485}{64} u^5+\frac{7371}{128} u^6\nonumber\\
&& +\frac{76545}{512} u^7+\frac{408969}{1024} u^8 
+\frac{17826237}{16384} u^9\nonumber\\&& +O(u^{10})  \, ,
\end{eqnarray}
and where $a_{1\rm{SF}}(u)$ is the first-order gravitational self-force (GSF) contribution to the EOB radial potential $A(u; \nu)$; see below.
The relation above, Eqs. (\ref{avshkk}) and (\ref{xi_exp}), is so simple that one can immediately read off the PN expansion of $h_{kk}(u)$ from that of $a_{1\rm{SF}}(u)$.
We shall therefore only give below the PN expansion of the EOB potential $a_{1\rm{SF}}(u)$ which has greater dynamical interest than $h_{kk}(u)$.

One should carefully distinguish between the GSF expansion of $A(u; \nu)$ (i.e., its expansion in powers of $\nu$) and its PN expansion (i.e., its expansion in powers of $u$, modulo some logarithms). 
To see the link between these two expansions it is useful to define the new function of two variables $a(u; \nu)$ such that
\beq
\label{A}
A(u; \nu)\equiv 1-2u +\nu a(u; \nu)\,,
\eeq
where we used the fact that, in the test mass limit $\nu \to 0$, $A(u; 0)$ reduces to the Schwarzschild radial potential $A^{\rm Schw}=1-2GM/(c^2r)=1-2u$. In terms of $a(u; \nu)$, the GSF expansion reads
\beq
\label{aSF}
a(u; \nu)=a_{1\rm{SF}}(u)+\nu a_{2\rm{SF}}(u)+O(\nu^2)\,,
\eeq 
while the PN expansion reads (up the 8.5PN order)
\begin{eqnarray}
\label{aPN}
a(u; \nu)&=&a_3(\nu)u^3 +a_4(\nu)u^4+a_5(\nu, \ln u)u^5\nonumber\\
&& +a_6(\nu, \ln u)u^6+a_{6.5}(\nu)u^{13/2}\nonumber\\
&& +a_7(\nu, \ln u)u^7+a_{7.5}(\nu)u^{15/2}\nonumber\\
&&+a_8(\nu, \ln u)u^8+a_{8.5}(\nu)u^{17/2}\nonumber\\
&& +a_9(\nu, \ln u)u^9
+a_{9.5}(\nu, \ln u)u^{19/2}\nonumber\\
&&+O_{\ln }(u^{10})\,,
\end{eqnarray}
where $O_{\ln }(u^{n})$ denotes some $O(u^n(\ln u)^p)$,  with an unspecified natural integer $p\ge 1$.
Note that the term $\sim a_n u^n$ in $a(u; \nu)$ corresponds to the $(n-1)$PN level, and that there is 
no contribution at 1PN: $a_2(\nu) \equiv 0$ (which is a special feature of the EOB formalism). 
Note also that we did not include the (a priori possible) argument $\ln u$ in $a_{6.5}, a_{7.5}$ and $a_{8.5}$ because we
have found that their $\nu \to 0$ limits contain no logarithmic contributions (and because previous analytical
derivations of logarithmic terms \cite{Damour:2009sm,Blanchet:2010zd} showed that they first appear in 
$\nu$-independent contributions).
These two expansions are linked by the fact that the $u$-expansion of the right-hand-side (r.h.s.) of Eq. (\ref{aSF}) must coincide with the $\nu$-expansion of the r.h.s. of Eq. (\ref{aPN}); e.g., the PN expansion of the
first-order GSF contribution $a_{1\rm{SF}}(u)$ is given by
\begin{eqnarray}
\label{aSFPN}
[a_{1\rm{SF}}(u)]^{PN}&=&a_3(0)u^3 +a_4(0)u^4+a_5(0, \ln u)u^5\nonumber\\
&& +a_6(0, \ln u)u^6+a_{6.5}(0)u^{13/2}\nonumber\\
&& +a_7(0, \ln u)u^7+a_{7.5}(0)u^{15/2}\nonumber\\
&&+a_8(0, \ln u)u^8+a_{8.5}(0)u^{17/2} \nonumber\\
&&+a_9(0, \ln u)u^9
+a_{9.5}(0, \ln u)u^{19/2}\nonumber\\
&&+O_{\ln }(u^{10})\,.
\end{eqnarray}
Up to now, the analytic knowledge of the PN expansion of $a(u; \nu)$ was the following. The 1PN and 2PN coefficients $a_2(\nu)=0$ and $a_3(\nu)=2$ were derived in 
\cite{Buonanno:1998gg}; the 3PN coefficient $a_4(\nu)=94/3-41\pi^2/32$ was derived in \cite{Damour:2000we}, and the 4PN coefficient
\begin{eqnarray}
\label{a5}
a_5(\nu , \ln u)&=&-\frac{4237}{60}+\frac{128}{5}\gamma+\frac{64}{5}\ln u +\frac{256}{5}\ln(2)\nonumber\\
&&+\frac{2275}{512}\pi^2 +\nu \left(-\frac{221}{6}+\frac{41}{32}\pi^2 \right)\,,
\end{eqnarray}
where $\gamma=0.577\ldots$ denotes Euler's constant, was derived in \cite{Bini:2013zaa}. 
Up to the 4PN level included the full $\nu$-dependence of $a(u;\nu)$ is known, and was found, as just recalled, to be extremely simple: independence on $\nu$ up to the 3PN level, and linearity in $\nu$ at the 4PN level.
Beyond the 4PN level, one generally does not know the $\nu$-dependence of $a(u; \nu)$, apart from the $\nu$-dependence of the logarithmic contribution to the 5PN level, namely \cite{Blanchet:2010zd,Tiec:2011ab,Barausse:2011dq}  
\beq
a_6(\nu , \ln u)=\left(-\frac{7004}{105}-\frac{144}{5}\nu  \right)\ln u +a_6^c(\nu)\,.
\eeq
On the other hand, the PN-expansion of the $\nu \to 0$ limit of $a(u; \nu)$ [i.e., the PN expansion of the first-order GSF contribution $a_{1\rm{SF}}(u)=a(u,\nu=0)$] was analytically determined up to the 6PN order in our previous work \cite{bdtd_beyond} with the following results
\begin{eqnarray}
\label{a6}
a_6(0, \ln u) &=&-\frac{1066621}{1575}-\frac{14008}{105}\gamma-\frac{7004}{105}\ln(u)\nonumber\\
&& -\frac{31736}{105}\ln(2)+\frac{246367}{3072}\pi^2\nonumber\\
&& +\frac{243}{7}\ln(3)\,, \\
\label{a65}
a_{6.5}(0) &=& +\frac{13696}{525}\pi\,,\\
\label{a7}
a_7(0, \ln u) &=& \frac{206740}{567}\ln(2)-\frac{2522}{405}\ln(u)-\frac{5044}{405}\gamma\nonumber\\
&& -\frac{1360201207}{907200}-\frac{4617}{14}\ln(3) \nonumber\\ 
&& -\frac{2800873}{262144}\pi^4 +\frac{608698367}{1769472}\pi^2\,.
\end{eqnarray}
The result Eq. (\ref{a65}) for $a_{6.5}(0)$  was independently analytically derived by Shah  {\it et al.}  \cite{Shah:2013uya}. 
The latter reference gave also a numerical-analytical derivation of the coefficient of $\ln u$ in $a_7$.
In addition Ref. \cite{Shah:2013uya} inferred from their high-accuracy numerical results plausible analytical expressions for several higher order coefficients related to $a_{10}^{\ln^2}$, $a_{10.5}^{\ln}$  and $a_{11}^{\ln^3}$ (see below).
 
The main result of this paper will be to extend the analytical calculation of the $a_n(0,\ln u)$'s up to the 8.5PN level, i.e., to determine five more terms beyond the ones listed above, namely $a_{7.5}(0)$, $a_8(0, \ln u)$, $a_{8.5}(0)$, $a_9(0, \ln u)$ and $a_{9.5}(0, \ln u)$. We shall find that our analytical results agree with the very accurate numerical determination of the PN expansion coefficients of
\beq
\label{Detw}
u^t_{1\rm SF}(u)=\frac1{2(1-3u)^{3/2}}h_{kk}^R(u) 
\eeq
recently obtained by Shah {\it et al.}  \cite{Shah:2013uya}. (The latter reference also provided  numerical-analytical derivations of the coefficients of several logarithmic contributions which agree with our corresponding fully analytic results.)

A second way to present our results consists of exhibiting the analytic expressions of the PN expansion coefficients of
Detweiler's original first-order redshift function, namely $u^t_{1\rm SF}(u)$ given by Eq. (\ref{Detw}).  We will do so
in Sec. III below, after having exhibited the analytic expressions of the $a_n$'s in Sec. II. 
Finally, a last way to present our results is in terms of the function relating the binding energy $E_B=H^{\rm tot}-Mc^2$ of a circular orbit to its orbital frequency $\Omega$. This alternative reformulation of our results will be presented in the Appendix.

\section{New analytical results for the EOB radial potential}

As mentioned above, our new results concern the coefficients of the PN expansion of the first-order GSF contribution $a_{1\rm SF}(u)$ to the EOB radial potential $A(u; \nu)$, Eqs. (\ref{A}) and (\ref{aSF}), between the 6.5PN level and the 8.5PN one. Namely, we found that the coefficients $a_{7.5}(0)$, $a_8(0, \ln u)$, $a_{8.5}(0)$, $a_9(0, \ln u)$ and $a_{9.5}(0, \ln u)$ in Eq. (\ref{aSFPN}) are given by
\begin{widetext}
\begin{eqnarray}
\label{anew75}
a_{7.5}(0)&=& -\frac{512501}{3675}\pi  \,,\\
\label{anew8}
a_8(0, \ln u)&=&  -\frac{187619320956191}{12224520000}+\frac{14667859963}{5457375}\gamma+\frac{19361011651}{5457375}\ln(2)\nonumber\\
&&+\frac{2048}{5}\zeta(3)-\frac{109568}{525}\gamma^2+\frac{1836927775597}{2477260800}\pi^2+\frac{14667859963}{10914750}\ln(u)-\frac{438272}{525}\ln^2(2)\nonumber\\
&& +\frac{3572343}{3520}\ln(3)+\frac{1953125}{19008}\ln(5)-\frac{438272}{525}\ln(2)\gamma-\frac{27392}{525}\ln^2(u)-\frac{109568}{525}\gamma\ln(u)\nonumber\\
&&-\frac{219136}{525}\ln(2)\ln(u)+\frac{830502449}{16777216}\pi^4\,,\\
\label{anew85}
a_{8.5}(0)&=&+\frac{70898413}{6548850}\pi  \,, \\
\label{anew9}
a_9(0, \ln u)&=&  \frac{3121123440903397043}{8899450560000}-\frac{23033337928985}{6442450944}\pi^4 -\frac{53276112149251}{92484403200}\pi^2-\frac{1198510638937}{198648450}\gamma\nonumber\\
&& -\frac{11647126988311}{993242250}\ln(2)-\frac{152128}{105}\zeta(3)+\frac{10894496}{11025}\gamma^2-\frac{1193425238617}{397296900}\ln(u)\nonumber\\
&&+\frac{322400}{63}\ln^2(2)-\frac{18954}{49}\ln^2(3)+\frac{325284577623}{71344000}\ln(3)-\frac{2283203125}{1482624}\ln(5)\nonumber\\
&&+\frac{17379776}{3675}\ln(2)\gamma-\frac{37908}{49}\gamma\ln(3)-\frac{37908}{49}\ln(2)\ln(3)+\frac{2723624}{11025}\ln^2(u)\nonumber\\
&&-\frac{18954}{49}\ln(u)\ln(3) +\frac{10894496}{11025}\gamma\ln(u)+\frac{8689888}{3675}\ln(2)\ln(u)\,,\\
\label{anew95}
a_{9.5}(0, \ln u)&=&\left(\frac{3008350528127363}{1048863816000}+\frac{219136}{1575}\pi^2-\frac{23447552}{55125} \gamma 
-\frac{46895104}{55125} \ln(2)-\frac{11723776}{55125} \ln(u)\right)\pi  
 \,.
\end{eqnarray}
\end{widetext}
Note that the transcendentality\footnote{The designation ``transcendentality'' is used when discussing multi-loop 
Feynman integrals to order special numbers (especially multi-zeta values) in terms of their ``level''.}
 of the coefficients $a_n$ increases with $n$. This was already noted in \cite{bdtd_beyond} up to $a_7$, i.e., up to the 6PN order. Here the transcendentality further increases (when considering separately the integer and half-integer values of $n$). In particular, we note that at the 7PN level ($a_8$) there appears (beyond transcendental numbers that entered previous levels, namely $\zeta(2)=\frac{\pi^2}{6}$, $\zeta(4)=\frac{\pi^4}{90}$, Euler's constant $\gamma$ and some logarithms), the value of the zeta function at 3, $\zeta(3)$, as well as the square of $\gamma$. The appearance of $\gamma^2$ is linked with the appearance of the square of $\ln u$ (which starts at the 7PN level; as was pointed out in \cite{Shah:2013uya}). Indeed, the work of Refs. \cite{bdtd_beyond,DJS2014} shows that $\gamma$ enters via tail logarithms of the type
\beq
\label{taillog}
\ln \left(2 |m| \frac{\Omega r}{c}e^\gamma  \right)=\frac12 \ln u +\gamma +\ln (2|m|)\,,
\eeq
where $m$ denotes the \lq\lq magnetic" index in a corresponding spherical-harmonics $(lm)$ decomposition, so that $-l\le m \le l$.
(Actually, depending on the parity of the relevant mode, one has either $|m|=l,l-2,l-4,\ldots$, or $|m|=l-1,l-3,\ldots$). The leading-order near-zone tail being quadrupolar ($l=2$, even parity), we expect that the first $\gamma^2$ will enter in the combination $(\ln u+2\gamma +4\ln 2)^2$. One indeed finds that, e.g., $a_8(0, \ln u)$ can be more simply written as
\begin{eqnarray}
\label{a8rewritten}
a_8(0, \ln u)&=& -\frac{27392}{525}(\ln u+2\gamma +4\ln 2)^2\nonumber\\
&& +\frac{14667859963}{10914750}(\ln u +2\gamma)\nonumber\\
&&+\frac{19361011651}{5457375} \ln 2 +\frac{3572343}{3520}\ln 3\nonumber\\
&& +\frac{1953125}{19008}\ln 5+\frac{1836927775597}{2477260800}\pi^2\nonumber\\
&& +\frac{830502449}{16777216}\pi^4\nonumber\\
&&
+\frac{2048}{5}\zeta(3)
-\frac{187619320956191}{12224520000}\,.
\end{eqnarray}

Let us also note that, at the 8.5 PN level, there starts to be a mixing between the factor $\pi^1$ associated with (the reactive part of) tail terms
and the transcendentals $\zeta(2)$, $\gamma$, $\ln 2$, which generates $\propto \pi^3$, $\pi \gamma$ and $\pi \ln 2$.

\section{New analytic results for the Detweiler redshift function}

In order to explicitly compare our analytic results to the recent (mainly numerical) results of \cite{Shah:2013uya} let us express our results in terms of the original first-order Detweiler function, Eq. (\ref{Detw}), that was computed in \cite{Shah:2013uya}. Using the same notation for the PN coefficients as the latter reference (which differs from the one we used in \cite{bdtd_beyond} by introducing minus signs associated with odd powers of $\ln u$), our new analytic results are encoded in the coefficients $\alpha_n$, $\beta_n$ and $\gamma_n$ (with $6.5\le n\le 8.5$) entering 
\begin{eqnarray}
u_1^t (u) &=&u_1^t \big|_{6PN} +\alpha_{6.5} u^{15/2}\nonumber\\
&& +
(\alpha_7 -\beta_7 \ln u +\gamma_7\ln^2 u)u^8\nonumber\\
&& +
\alpha_{7.5} u^{17/2}+(\alpha_8 -\beta_8 \ln u +\gamma_8\ln^2 u)u^9\nonumber\\
&&+ (\alpha_{8.5} -\beta_{8.5} \ln u)u^{19/2}\,.
\end{eqnarray} 
From Eqs. (\ref{anew75})--(\ref{anew95}) above, we find
\begin{widetext}
\begin{eqnarray}
\alpha_{6.5}&=& \frac{81077}{3675}\pi\,,\nonumber\\
\alpha_7 &=& -\frac{10327445038}{5457375}\gamma-\frac{16983588526}{5457375}\ln(2)-\frac{2048}{5}\zeta(3)
-\frac{2873961}{24640}\ln(3)-\frac{1953125}{19008}\ln(5)+\frac{438272}{525}\ln(2)\gamma\nonumber\\
&&-\frac{23851025}{16777216}\pi^4-\frac{9041721471697}{2477260800}\pi^2+\frac{109568}{525}\gamma^2
+\frac{438272}{525}\ln^2(2)+\frac{12624956532163}{382016250}\,,\nonumber\\
\beta_7 &=& -\left(\frac{109568}{525}\gamma+\frac{219136}{525}\ln(2)-\frac{5163722519}{5457375}\right)\,, \nonumber\\
\gamma_7 &=& \frac{27392}{525}\,,\nonumber\\
\alpha_{7.5}&=& +\frac{82561159}{467775}\pi\,, \nonumber\\
\end{eqnarray}
\begin{eqnarray}
\alpha_8 &=&  -\frac{1526970297506}{496621125}\gamma-\frac{1363551923554}{496621125}\ln(2)
-\frac{41408}{105}\zeta(3)-\frac{2201898578589}{392392000}\ln(3)\nonumber\\
&&+\frac{798828125}{741312}\ln(5)-\frac{3574208}{3675}\ln(2)\gamma+\frac{37908}{49}\gamma\ln(3)
+\frac{37908}{49}\ln(2)\ln(3)+\frac{22759807747673}{6442450944}\pi^4\nonumber\\
&&-\frac{246847155756529}{18496880640}\pi^2-\frac{108064}{2205}\gamma^2
-\frac{2143328}{1575}\ln^2(2)+\frac{18954}{49}\ln^2(3)
-\frac{7516581717416867}{34763478750}\,,\nonumber\\
\beta_8 &=& -\left(\frac{18954}{49}\ln(3)-\frac{108064}{2205}\gamma-\frac{1787104}{3675}\ln(2)
 -\frac{769841899153}{496621125}\right)\,,\nonumber\\
\gamma_8 &=& -\frac{27016}{2205}\,,\nonumber\\
\alpha_{8.5} &=& \left(-\frac{2207224641326123}{1048863816000}-\frac{219136}{1575}\pi^2
 +\frac{23447552}{55125} \gamma+\frac{46895104}{55125} \ln(2)\right)\pi\,,\nonumber\\
\beta_{8.5} &=& -\frac{11723776}{55125} \pi  \,.
\end{eqnarray}
\end{widetext}
The analytical expressions for $\alpha_{6.5}$, $\gamma_7$, $\alpha_{7.5}$, $\gamma_8$ and $\beta_{8.5}$ derived here from our  results coincide with the corresponding analytic expressions inferred by Shah et al. from their numerical results. On the other hand, the analytic expressions of the other coefficients (which have a high transcendentality structure), i.e., $\alpha_7$, $\beta_7$, $\alpha_8$, $\beta_8$ and $\alpha_{8.5}$ have been obtained here for the fist time.
We have checked that their numerical values  
\begin{eqnarray}
\alpha_7 &=& -6343.874453\ldots \nonumber\\
\beta_7 &=&  536.405212\ldots \nonumber\\
\alpha_8 &=&-11903.472947\ldots \nonumber\\
\beta_8 &=& 1490.555085\ldots \nonumber\\
\alpha_{8.5} &=& -8301.373708\ldots\,,
\end{eqnarray}
agree with the numerical results obtained in \cite{Shah:2013uya} (to the accuracy given there).

\section{Numerical values of higher-order PN expansion coefficients of the EOB radial potential}

Shah et al. \cite{Shah:2013uya} succeded in numerically computing the expansion coefficients $\alpha_n$, $\beta_n$ and $\gamma_n$ of $u_{1\rm SF}^t$ up to the 10.5PN order.
Moreover, they also inferred from their numerical results the (probable) analytic expression of some specific terms (having a minimal transcendentality structure). Using the relations (\ref{Detw}) and (\ref{avshkk}), one can deduce the numerical values of the corresponding
  higher-order coefficients of the PN expansion of the first-order GSF EOB radial potential, $a_{1\rm SF}(u)$. We find
\begin{eqnarray}
a_{1\rm SF}(u)&=& a_{1\rm SF}^{\le 8.5\rm PN}(u)+a_{10}(0,\ln u)u^{10}+a_{10.5}(0,\ln u)u^{21/2}\nonumber\\
&&+a_{11}(0,\ln u)u^{11}+a_{11.5}(0,\ln u)u^{23/2}\nonumber\\
&&+O_{\ln}(u^{12})
\end{eqnarray}
where
\beq
a_n(0,\ln u)=a_n^c+a_n^{\ln} \ln u +a_n^{\ln^2}\ln^2 u +a_n^{\ln^3} \ln^3 u\,,
\eeq 
with
\begin{eqnarray}
a_{10}^c&=&
-\frac{539189410745499497}{494413920000}-\frac{272499972037}{35315280}\gamma\nonumber\\
&&-\frac{13728}{35}\zeta(3)-\frac{78745197816729}{3139136000}\ln(3)\nonumber\\
&&+\frac{3423828125}{658944}\ln(5)+\frac{68222591991915}{4294967296}\pi^4\nonumber\\
&&-\frac{953787855261929}{20552089600}\pi^2-\frac{1822982697461}{882882000}\ln(2)\nonumber\\
&&-\frac{8812704}{1225}\ln(2)\gamma+\frac{170586}{49}\gamma\ln(3)\nonumber\\
&&+\frac{170586}{49}\ln(2)\ln(3)-\frac{312944}{35}\ln(2)^2\nonumber\\
&&-\frac{1133008}{1225}\gamma^2+\frac{85293}{49}\ln(3)^2-\alpha_9\nonumber\\
&=&4845.8705570194441773473934215798222\,,\nonumber\\
a_{10}^{\ln}&=&  -\frac{276568292293}{70630560}-\frac{1133008}{1225}\gamma\nonumber\\
&&-\frac{4406352}{1225}\ln(2)+\frac{85293}{49}\ln(3)+\beta_9\nonumber\\
&=&-8207.44191517196106149591367198537806\,,\nonumber\\
a_{10}^{\ln^2}&=&\frac{50176712}{280665}= 178.7779452\ldots
\end{eqnarray}
\begin{eqnarray}
a_{10.5}^c&=& -\frac{2343200017302563}{233080848000}\pi-\frac{109568}{175}\pi^3\nonumber\\
&&+\frac{11723776}{6125}\pi\gamma+\frac{23447552}{6125}\pi\ln(2)-\alpha_{9.5}\nonumber\\
&=&-28324.307465213628065671194\,,\nonumber\\
a_{10.5}^{\ln}&=&\frac{1207083334}{1157625}\pi= 3275.813960\ldots
\end{eqnarray}
\begin{eqnarray}
a_{11}^c&=&-\alpha_{10}+\frac92\alpha_9-\frac{255879}{98}\gamma\ln(3)\nonumber\\
&&+\frac{1566004413023}{117717600}\gamma+\frac{2809348340741}{196196000}\ln(2)\nonumber\\
&&+\frac{2354788692879445429}{3515832320000}-\frac{1521484375}{439296}\ln(5)\nonumber\\
&&+\frac{111444}{35}\ln(2)^2-\frac{255879}{196}\ln(3)^2\nonumber\\
&&+\frac{2295288}{1225}\ln(2)\gamma+\frac{4335785656914767}{82208358400}\pi^2\nonumber\\
&&-\frac{255879}{98}\ln(2)\ln(3)+\frac{118690107424281}{6278272000}\ln(3)\nonumber\\
&&+\frac{70776}{35}\zeta(3)-\frac{228804}{1225}\gamma^2\nonumber\\
&&-\frac{205145528573025}{17179869184}\pi^4\nonumber\\
&=&135603.46094278\,,\nonumber\\
a_{11}^{\ln}&=& \frac{1147644}{1225}\ln(2)+\beta_{10}-\frac92\beta_9-\frac{255879}{196}\ln(3)\nonumber\\
&&-\frac{228804}{1225}\gamma+\frac{1576175213663}{235435200} \nonumber\\
&=& 12739.961666212\,,\nonumber\\
a_{11}^{\ln^2}&=& -\frac{412951489}{218295}-\gamma_{10}\nonumber\\
&=& -3997.64018516794\,,\nonumber\\
a_{11}^{\ln^3}&=&\frac{23447552}{165375}= 141.7841391\ldots
\end{eqnarray}
\begin{eqnarray}
a_{11.5}^c&=& 
-\frac{17585664}{6125}\pi\ln(2)+\frac{9}{2}\alpha_{9.5}-\alpha_{10.5}\nonumber\\
&&+\frac{708244844441941}{103591488000}\pi+\frac{82176}{175}\pi^3\nonumber\\
&&-\frac{8792832}{6125}\pi\gamma\nonumber\\ 
&=& -82057.0\,,\nonumber\\
a_{11.5}^{\ln} &=& -\frac{141917987}{128625}\pi+\beta_{10.5}= 1546.9\,.
\end{eqnarray}

Here, the number of digits quoted for the $a_n$'s is that corresponding to the numerical accuracy on the $\alpha_n$'s etc. obtained in \cite{Shah:2013uya}.
It would be interesting to see if the numerical values of these coefficients could be improved by using in the fits done in Ref. \cite{Shah:2013uya} the exact analytical values of the coefficients below the 9PN order given by our analytical results above. This additional analytic knowledge would indeed significantly decrease the number of fitted parameters  and thereby allow a better decorrelation of the \lq\lq signal" associated with the higher-order ones.
It might allow one to extract even higher PN coefficients from the numerical results of \cite{Shah:2013uya}.

\section{Comparison to the previous numerical determination of the EOB function $a_{1\rm SF}(u)$ in the strong-field domain}

Akcay {\it et al} \cite{Akcay:2012ea} succeded in numerically computing the global behavior of the function $a_{1\rm SF}(u)$ over the full strong-field interval, $0\le u \le 1/3$, where it can be probed by studying the sequence of circular orbits (see also Refs. \cite{Detweiler:2008ft,Blanchet:2009sd,Blanchet:2010zd} for previous numerical results). Let us study to what extent the knowledge of the higher-order PN expansion of   $a_{1\rm SF}(u)$ (which, a priori, only encodes information about the weak-field regime, $u\ll 1$) can represent some of the strong-field features of that function.

A first important strong-field feature of $a_{1\rm SF}(u)$ is its behavior around the last stable (circular) orbit (LSO). In particular, a benchmark quantity is the first-order GSF shift in the LSO orbital frequency  \cite{Barack:2009ey}. It is measured by the coefficient $c_\Omega$ such that
\beq
(m_1+m_2)\Omega_{\rm LSO}=6^{-3/2}[1+c_\Omega \nu +O(\nu^2)]\,.
\eeq
It was found in \cite{Damour:2009sm} that $c_\Omega$ can be expressed as follows in terms of the first two derivatives of the EOB potential $a_{1\rm SF}(u)$ at $u=1/6$:
\beq
\label{COmega}
c_\Omega=\frac32 \tilde a(\frac16) +1-\frac{2\sqrt{2}}{3}\,,
\eeq 
where
\beq
\label{atilde}
 \tilde a(\frac16)=  a_{1\rm SF}(\frac16) + \frac16 a'_{1\rm SF}(\frac16) + \frac1{18} a''_{1\rm SF}(\frac16)  \, .
\eeq
The numerical value of $c_\Omega$ was first obtained in \cite{Barack:2009ey} and then refined in later works. The best current estimate \cite{Akcay:2012ea} is
\beq
\label{COmeganum}
c_\Omega^{\rm num}=1.251015464(46)\,,
\eeq
corresponding to 
\beq
 \tilde a^{\rm num}(\frac16)=0.795883004(30)\,.
\eeq
In the second column of Table I  we list the values of $c_\Omega$ obtained by inserting in Eqs. (\ref{COmega}), (\ref{atilde}) the successive PN approximants of $a_{1\rm SF}(u)$ (starting with $a_{1\rm SF}^{\rm 2PN}(u)=2u^3$ and going up to the 10.5PN approximant deduced in the previous section from the results of \cite{Shah:2013uya}). This Table does not exhibit a clear convergence towards the accurate result (\ref{COmeganum}). In fact, while the 7PN and 7.5PN results suggest some convergence towards the exact value, the results significantly worsen between the 8PN and the 10PN levels. Even the highest known level (10.5PN) fares less well  than the 7 or 7.5 PN levels. This is another example of the lack of convergence of the PN expansion in the strong-field domain.

\begin{table}[t]
  \caption{\label{tab:pnorder}}
  \begin{center}
    \begin{ruledtabular}
      \begin{tabular}{lcc}
 PN order  & $c_{\Omega}$ & $\hat a_E(1/3)$\\
\hline
2&  .2794131806 & 1 \cr
3& .9066691736 & 4.281\cr
4& .9586305806 & 5.311\cr
5&.8779575416 & 4.158\cr
5.5&.9629380916 & 5.034\cr
6& 1.331486606 & 9.128\cr
6.5& 1.229896832 & 7.713\cr
7&  1.253181340 & 9.505\cr 
7.5& 1.254880160 & 9.273 \cr
8&  1.061895577 &2.765 \cr
8.5& 1.184747405 &7.259 \cr\hline 
9 (Ref. \cite{Shah:2013uya}) & 1.271563518& 8.810 \cr    
9.5 (Ref. \cite{Shah:2013uya}) &1.199854811& 5.387\cr  
10 (Ref. \cite{Shah:2013uya})& 1.299914640& 14.183 \cr    
10.5 (Ref. \cite{Shah:2013uya}) & 1.263663130& 9.791 \cr\hline 
Numerical (Ref. \cite{Akcay:2012ea}) &1.251015464(46) & 10.19(3)\cr
   \end{tabular}
  \end{ruledtabular}
\end{center}
\end{table}

\begin{figure}
\label{fig:1}
\begin{center}
\includegraphics[scale=0.35]{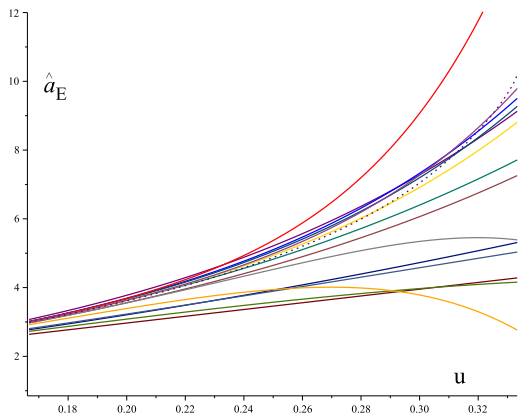}
\end{center}
\caption{The plot compares, on the interval $\frac16 \le u\le \frac13$, the successive  PN approximants of  the quantity $\hat a_E(u)=a(u)/(2u^3 E(u))$ with $E(u)=(1-2u)/\sqrt{1-3u}$, to its ``exact'' numerical value. There are fifteen curves  in the figure:  fourteen PN approximants (from 3PN to 10.5 PN), and the fit model \#14 of numerical relativity data in Ref. \cite{Akcay:2012ea}. The curves are distinguishable from their values at $u=\frac13$, where   they are ordered from bottom to top as follows:  8PN, 5PN, 3PN, 5.5PN, 4PN, 9.5 PN, 8.5PN, 6.5PN, 9PN, 6PN, 7.5PN, 7PN, 10.5PN, model \#14 (dotted curve), 10PN.}
\end{figure}

To further study the convergence properties of the PN expansion let us compare the successive PN approximants of $a_{1{\rm SF}}(u)$ to the numerically-determined 
global computation of $a_{1{\rm SF}}(u)$. As it was found in \cite{Akcay:2012ea} that the function $a_{1{\rm SF}}(u)$ diverges proportionally to the test-particle energy
\beq
\label{E}
E(u):= \frac{1-2u}{\sqrt{1-3u}}
\eeq
as $u$ approaches the light ring ($u\to (1/3)^-$), a meaningful PN/numerics comparison must factor out the divergent factor (\ref{E}). (From Weierstrass' theorem, one could hope that the {\it continuous}, and therefore bounded, function 
$a_{1{\rm SF}}(u)/E(u)$ on the {\it closed} interval $0\leq u\leq \frac13$ will be approximable by its PN expansion, which is, modulo some logarithms, a polynomial).
It is also convenient to factor
out the leading order PN approximation  $a_{1{\rm SF}}^{\rm 2PN}(u)=2u^3$, and therefore to work with the following doubly rescaled $a$-potential \cite{Akcay:2012ea}
\beq
\label{aE}
\hat a_E(u)=\frac{a_{1{\rm SF}}(u)}{2u^3E(u)}\,.
\eeq
In Fig. 1   
we compare the successive PN approximants of $\hat a_E(u)$ 
[starting with  $a_{1{\rm SF}}^{\rm 3PN}(u)=1+(97/6-41\pi^2/64)u$] to the numerical value of $\hat a_E(u)$ (as conventionally encoded in the accurate fit \# 14 in \cite{Akcay:2012ea}). 
This figure clearly shows that, even after the factorization of the divergent factor (\ref{E}) that is known to be present in $a_{1{\rm SF}}(u)$, the sequence of PN approximants does not converge globally (i.e., in the full interval $0\le u\le 1/3$) towards $\hat a_E(u)$.
Note in particular the huge scatter of values reached by the various $a_{1{\rm SF}}^{\rm nPN}(u)$'s at $u=1/3$. While the correct numerical value is 
$\hat a_E(1/3)=10.19(3)$, the corresponding PN approximants range between  $\hat a_E^{\rm 8PN}(1/3)=2.765$ to  $\hat a_E^{\rm 10PN}(1/3)=14.183$. 
The numerical values of $\hat a_E^{n\rm PN}(1/3)$ are listed in the third column of table I.
Even if we focus on the last PN approximants, the scatter remains very large, and non monotonic.
The fact that the last PN approximant (10.5PN) turns out to be rather close to the exact value is probably coincidental (or, possibly helped by the fact that the fits done by Shah et al.  \cite{Shah:2013uya} have absorbed in the 10.5PN level the \lq\lq signal" contained in the remaining, non fitted, PN terms).
Even if we were to restrict the interval to, say, $0\le u\le 1/6$ (i.e., for radii $r=GM/(c^2 u)$ above the LSO) a close look at the various curves shows that there is no monotonicity in the way the successive PN approximants approach the exact result.

\section{Conclusions}
\label{section:concluding}

Let us summarize our results:
we have studied several functions characterizing in a {\it gauge-invariant} way the energetics of binary systems in the limit $m_1\ll m_2$. At linear order in the symmetric mass ratio $\nu=m_1m_2/(m_1+m_2)^2$ these functions $h_{kk}^R(u)$, $a_{1\rm SF}(u)$ and $u^t_{1\rm SF}(u)$ are linked by simple relations, see Eqs. (\ref{avshkk}) and (\ref{Detw}). We focussed on the function  $a_{1\rm SF}(u)$ which encodes the $\nu$-linear correction to the EOB function $A(u;\nu)$, see Eq. (\ref{A}). Indeed, within the EOB formalism, the function $A(u;\nu)=1-2u+\nu a_{1\rm SF}(u)+O(\nu^2)$ plays a central role because it parametrizes the time-time component of the EOB effective metric, 
\beq
ds^2=-A(r; \nu)dt^2+B(r;\nu)dr^2 +r^2(d\theta^2+\sin^2\theta d\phi^2)\,,
\eeq
and thereby plays (similarly to its well-known test-mass limit $A(r; 0)=1-2GM/(c^2r)=1-2u$) the role of main radial potential describing the two-body gravitational interaction.

Using techniques that were expounded in detail in Refs. \cite{Bini:2013zaa,bdtd_beyond} we extended here our previous 6PN-accurate results by deriving, for the first time, the {\it analytic} expression of the PN expansion of $a_{1\rm SF}(u)$ up to the 8.5 PN level included. See Eqs. (\ref{anew75})--(\ref{anew95}). We compared our results with the (mainly numerical) results of \cite{Shah:2013uya} and found perfect agreement. We hope that the new knowledge brought by our results will allow one to extract more information from the very accurate numerical simulations of   \cite{Shah:2013uya}. We then transcribed the results of  \cite{Shah:2013uya} going beyond the 8.5PN level  into the numerical knowledge of several higher-order contributions to the first GSF-order EOB potential $a_{1\rm SF}(u)$. (For a few terms, Ref. \cite{Shah:2013uya} also provided plausible analytic expressions).

Armed with the knowledge of the PN expansion of $a_{1\rm SF}(u)$ up to the 10.5 PN level, we studied the convergence of the PN expansion as the expansion parameter $u=GM/(c^2r)$ leaves the weak-field domain $u\ll 1$ to enter the strong field regime $u=O(1)$. This study involved a comparison of the accurate knowledge of the strong-field behavior of the function $a_{1\rm SF}(u)$ which was recently obtained \cite{Akcay:2012ea} to some of its PN approximants. First, we considered the parameter
$c_\Omega$ measuring the first-order GSF shift of the orbital frequency of the last stable (circular) orbit. Second, we considered the global behavior, on the interval $0\le u \le 1/3$, of a regularized version of the function  $a_{1\rm SF}(u)$ introduced in \cite{Akcay:2012ea} [Indeed, the latter reference found that $a_{1\rm SF}(u)$ was singular at $u=1/3$, i.e., at the light ring, so that only suitably regularized (bounded) versions of $a_{1\rm SF}(u)$ can a priori be expected  to be potentially representable as a PN expansion (which is, essentially, a polynomial in $u$)]. However, our results show that the PN expansion, even considered up to the 10.5PN level, fails to provide an accurate representation of $a_{1\rm SF}(u)$ (with or without regularization at the light ring). Even, if we focus on the semi-strong-field behavior around $u=1/6$ (LSO), our study shows an, at best, erratic convergence of the PN expansion towards exact, numerical results  (However, a few PN approximants seem, coincidentally, to provide a reasonably accurate value for the LSO frequency shift $c_\Omega$; notably the 7PN and 7.5PN approximants). To complete our results, we present, in the Appendix, the 8.5PN accurate expansion of the function linking the binding energy of a circular orbit to its orbital frequency.

To conclude, let us emphasize that, by using the method we presented in \cite{Bini:2013zaa,bdtd_beyond}, our analytic derivation could be extended, with some limited additional effort, to higher PN orders. Such an extension might be interesting both as a test-bed for accurate numerical computation, and also as a case-study of the increase in the transcendentality of the coefficients entering the computation of multi-loop Feynman integrals. Indeed, our computations, when viewed in terms of gravitational perturbation theory around a flat spacetime, involve higher and higher Feynman-like \lq\lq loop integrals" \cite{Damour:1995kt,Damour:2001bu}.
The study of the special numbers entering multi-loop integrals has recently become of interest in mathematical physics \cite{Broadhurst:1995km,Brown:2009ta}. From this point of view, an interesting outcome of our result is the first appearance of $\zeta(3)$ at the 7PN order, corresponding to 7 loops in a flat space expansion (though it is here computed from a one-loop effect around a {\it curved} manifold).

Let us note a potentially useful practical consequence of an a priori analytic knowledge of the transcendental
content of some PN coefficient: it can allow one to (plausibly) infer its full analytic structure from a sufficiently accurate
numerical computation. For instance, if we only know that, after separation of the terms $\propto (\ln u+2\gamma +4\ln 2)^2$
and $\propto (\ln u+2\gamma )^1$ (with rational coefficients), the remaining 7PN coefficient $a_8^{\rm no \, log}$, last seven terms in Eq. (\ref{a8rewritten}),  
is a {\it rational} linear combination of $1, \ln 2, \ln 3, \ln 5$ [as follows from Eq. (\ref{taillog}], $\pi^2$, $\pi^4$ and 
$\zeta(3)$, we have found (via some numerical tests with existing integer-relations algorithms) that the numerical knowledge
of $\geq 162$ digits of   $a_8^{\rm no \, log}$ is sufficient to infer the seven rational coefficients entering the last
lines of  Eq. (\ref{a8rewritten}).
On the other hand, if we know the rational value of the coefficient of $\ln 5$ (which is rather easy to compute from the analytic understanding of near-zone tail effects), it is enough to know 141 digits to infer the remaining six rational coefficients. In addition, if we use the knowledge of the PN-expanded Regge-Wheeler-Zerilli solutions (which are much easier to compute that the hypergeometric-expansion ones, see Ref. \cite{bdtd_beyond}), one can also a priori compute the rational coefficients of $\pi^2$ and $\pi^4$ (which come from summing infinite series in $l$). In that case, it is enough to know $\ge 76$ digits to infer the remaining four rational coefficients. This study shows how crucial any (even partial) analytic knowledge can be in allowing one to infer exact results from numerical ones. Indeed, though transcendentals such as $\pi^2$, $\pi^4$, $\ln 2$, ..., are linearly independent over the rationals, they are approximately dependent to a surprisingly high accuracy.

Another conclusion of our work is that it would very valuable to be able to compute {\it gauge-invariant} GSF quantities at the {\it second order} in $\nu$.
Indeed, the recent progress (including the present paper) on the knowledge of the {\it first order} GSF contribution to the main EOB radial potential is sufficient for all practical purposes, especially in view of the global numerical knowledge of $a_{1\rm SF}(u)$ and of the poor convergence of its PN expansion.
By contrast, little is known about its second order GSF contribution,  apart from: its leading-order 4PN contribution \cite{bdtd_beyond} $\nu a_5^{c1}u^5$, Eq. (\ref{a5}); the indication of its singularity structure at the light ring \cite{Akcay:2012ea}; and the $\nu$-dependence of $A(u; \nu)$ inferred from comparisons to full numerical-relativity simulations  \cite{Damour:2012ky}.

\appendix

\section{The binding energy function}
As discussed in \cite{bdtd_beyond} (see Sec. IIIA there) , the EOB formalism provides a simple way to derive the functional link between the binding energy
$E_B\equiv H^{\rm tot}-Mc^2$ and the 
 dimensionless orbital frequency parameter $x=(M\Omega)^{2/3}$. 
Here, we shall work with the rescaled, and Newton-factorized, binding energy, i.e., the function $\hat e(x; \nu)$ defined by 
\beq
\hat e(x; \nu)=\frac{E_B (x;\nu)}{(-\frac12 \mu c^2 x)} \,,
\eeq
which has the form
\begin{eqnarray}
\label{eq10}
\hat e(x; \nu) &= & 1+ e_{\rm 1PN} (\nu)x 
+ e_{\rm 2PN} (\nu)x^2 +\ldots   \, .
\end{eqnarray}
Following the same notation and conventions used in \cite{bdtd_beyond}, we can analytically compute the next terms, from the $6.5$ PN approximation up to the $8.5$ one; these are given by
\begin{eqnarray}
e_{\rm 6.5PN} (\nu)&=& e_{\rm 6.5PN}^{c}\nonumber\\
e_{\rm 7PN} (\nu, \ln x)&=&   e_{\rm 7PN}^{c} +e_{\rm 7PN}^{\ln}\ln x +e_{\rm 7PN}^{\ln^2 }\ln^2 x\nonumber\\
e_{\rm 7.5PN} (\nu)&=& e_{\rm 7.5PN}^{c} \nonumber\\
e_{\rm 8PN} (\nu, \ln x)&=&  e_{\rm 8PN}^{c} +e_{\rm 8PN}^{\ln}\ln x +e_{\rm 8PN}^{\ln^2 }\ln^2 x \nonumber\\  
e_{\rm 8.5PN} (\nu, \ln x)&=&  e_{\rm 8.5PN}^{c} +e_{\rm 8.5PN}^{\ln}\ln x\,.
\end{eqnarray}
As emphasized in \cite{Bini:2013zaa,bdtd_beyond}  the PN expansion coefficients  $e_{n {\rm PN}}(\nu, \ln x)$
have a more complicated structure than the coefficients $a_n(\nu, \ln u)$ entering the PN expansion of the EOB
potential $A(u; \nu)$. The $\nu$-dependence of the $a_n$'s is much more restricted. Let us, without loss of generality,
write the PN expansion, Eq. (\ref{A}), of $A(u;\nu)$ as
\begin{eqnarray}
A(u; \nu ) 
&=& 1-2u +\nu a_3(\nu) u^3 + \nu a_4(\nu) u^4 \nonumber\\
&&+ \nu (a_5^c(\nu)+a_5^{\ln} (\nu)\ln u)u^5\nonumber\\
&&+ \nu (a_6^c(\nu)+a_6^{\ln} (\nu)\ln u)u^6+\nu  a_{6.5}^c(\nu) u^{13/2}\nonumber\\
&& +\nu (a_7^c(\nu)+a_7^{\ln} (\nu)\ln u)u^7+ \nu  a_{7.5}^c(\nu) u^{15/2}\nonumber\\
&& + \nu (a_8^c(\nu)+a_8^{\ln} (\nu)\ln u + a_8^{\ln^2} (\nu)\ln^2 u )u^8\nonumber\\
&& +
\nu ( a_{8.5}^c(\nu)+ +a_{8.5}^{\ln} (\nu)\ln u) u^{17/2}+\ldots
\end{eqnarray}
where the coefficients which are only known at the linear order in $\nu$ are parametrized by expressions of the type
\begin{eqnarray}
a_k^c(\nu) &=&a_k^c(0) + \nu a_k^{c1}(\nu)\nonumber\\
a_k^{\ln}(\nu) &=& a_k^{\ln}(0) + \nu a_k^{1 \ln}(\nu)\nonumber\\
a_k^{\ln^2}(\nu) &=& a_k^{\ln^2}(0) + \nu a_k^{1\ln^2}(\nu)\,,
\end{eqnarray}
for various $k$. 
Note in passing that, according to this notation, 
we can then write the GSF
expansion of $a(u,\nu)$, Eq. (\ref{aSF}), as
\begin{eqnarray}
\label{new_notation}
a_{1\rm{SF}}(u)&=&
\sum_{k\ge 3}a_k^c(0) u^k + \ln u\sum_{k\ge 5} a_k^{\ln} (0) u^k\nonumber\\
&& + \ln^2 u\sum_{k\ge 8} a_k^{\ln^2} (0) u^k\nonumber\\
&& +\sum_{k\ge 13}a_{k/2}^c(0)u^{k/2}+\ldots\nonumber\\
a_{2\rm{SF}}(u)&=&
\sum_{k\ge 5}a_k^{c1}(0) u^k + \ln u\sum_{k\ge 6} a_k^{1\ln} (0) u^k\nonumber\\
&& + \ln^2 u\sum_{k\ge 8} a_k^{1\ln^2} (0) u^k\nonumber\\
&& +\sum_{k\ge 13}a_{k/2}^{c1}(0)u^{k/2}+\ldots\,,
\end{eqnarray}

A straightforward calculation  then yields (suppressing, for brevity, the indication that the coefficients
$a^{c1}_n,  a^{1 \, \ln}_n,  a^{1 \, \ln^2}_n, \cdots$  depend on $\nu$)
\begin{widetext}
\begin{eqnarray}
e_{\rm 6.5PN}^{c}&=&   -\frac{20}{3}\nu^3 a^{c1}_{6.5}   +\left(-\frac{54784}{315}\pi+6 a^{c1}_{6.5} +4 a^{c1}_{7.5} \right)\nu^2-\frac{1474772}{3675}\nu\pi  
\end{eqnarray}
\begin{eqnarray}
e_{\rm 7PN}^{c}&=&   -\frac{12196899}{16384}-\frac{55913}{322486272}\nu^7-\frac{79079}{8957952}\nu^6+\left(-\frac{533533}{2985984}\pi^2
+\frac{78527813}{17915904}\right)\nu^5\nonumber\\
&&+\left(\frac{70915407929}{29859840}-\frac{185545243}{1990656}\pi^2+\frac{91}{24} a^{c1}_6 -\frac{10192}{1215}\ln(2)-\frac{5096}{1215}\gamma\right)\nu^4\nonumber\\
&&+\left(-\frac{32695}{9}\ln(2)-\frac{8}{3} a^{1\ln}_7 +\frac{1612460369}{663552}\pi^2-\frac{5091749}{110592}\pi^4-\frac{143}{18} a^{c1}_7 +\frac{455}{24} a^{c1}_6 +\frac{1053}{8}\ln(3)\right. \nonumber\\
&& \left. -\frac{20724677852507}{696729600}-\frac{15743}{9}\gamma\right)\nu^3\nonumber\\
&&+\left(\frac{117}{8} a^{c1}_6 +\frac{183573}{56}\ln(3)+a^{1\ln}_7 +\frac{13}{3} a^{c1}_8 +\frac{13}{2} a^{c1}_7 +\frac{71241703}{4718592}\pi^4-\frac{26212271058361}{1045094400}\right. \nonumber\\
&& \left. +\frac{80950181}{5103}\ln(2)+\frac{2}{3} a^{1\ln}_8 +\frac{35774791}{3645}\gamma-\frac{8528}{9}\pi^2\ln(2)-\frac{4264}{9}\pi^2\gamma+\frac{240020219947}{159252480}\pi^2\right)\nu^2\nonumber\\
&&+\left(-\frac{1424384}{1575}\gamma^2+\frac{7301042333}{50331648}\pi^4-\frac{5697536}{1575}\ln(2)^2+\frac{25390625}{57024}\ln(5)-\frac{5697536}{1575}\ln(2)\gamma \right. \nonumber\\
&& \left. +\frac{49507671791461}{7431782400}\pi^2+\frac{244217572198}{16372125}\ln(2)+\frac{26624}{15}\zeta(3)+\frac{170459308774}{16372125}\gamma+\frac{68052231}{24640}\ln(3)\right. \nonumber\\
&& \left. -\frac{263924477296340551}{3129477120000}\right)\nu\nonumber\\
e_{\rm 7PN}^{\ln}&=&  -\frac{135226}{1215}\nu^4+\left(-\frac{143}{18}a^{1\ln}_7 -\frac{25571}{18}\right)\nu^3+
\left(\frac{13}{2} a^{1\ln}_7 +\frac{4}{3} a^{1\ln^2 }_{8} +\frac{32704243}{7290}-\frac{2132}{9}\pi^2+\frac{13}{3} a^{1\ln}_8 \right)\nu^2\nonumber\\
&&+\left(\frac{85229654387}{16372125}-\frac{2848768}{1575}\ln(2)-\frac{1424384}{1575}\gamma\right)\nu \nonumber\\
e_{\rm 7PN}^{\ln^2 }&=& -\frac{356096}{1575}\nu+\frac{13}{3}\nu^2 a_8^{1 \ln^2} \ 
\end{eqnarray}
\begin{eqnarray}
e_{\rm 7.5PN}^{c}&=&  \frac{2960788}{1575}\nu^2\pi+\frac{133}{6}\nu^3 a^{c1}_{6.5} -\frac{722418584}{1403325}\nu\pi+7\nu^2 a^{c1}_{7.5} \nonumber\\
&&+\frac{54784}{405}\nu^3\pi+\frac{140}{27}\nu^4 a^{c1}_{6.5} -\frac{28}{3}\nu^3 a^{c1}_{7.5} +\frac{63}{4}\nu^2 a^{c1}_{6.5} +\frac{14}{3}\nu^2 a^{c1}_{8.5}   
\end{eqnarray}
\begin{eqnarray}
e_{\rm 8PN}^{c}&=& -\frac{70366725}{32768}-\frac{5}{32768}\nu^8-\frac{65}{8192}\nu^7+\left(-\frac{2255}{24576}\pi^2+\frac{323935}{147456}\right)\nu^6\nonumber\\
&&+\left(-\frac{14}{3}\ln(2)-\frac{872160733}{442368}-\frac{15}{16} a^{c1}_6 -\frac{7}{3}\gamma+\frac{14764015}{196608}\pi^2\right)\nu^5\nonumber\\
&&+\left(-\frac{3645}{112}\ln(3)-\frac{547845815}{98304}\pi^2-\frac{625}{8} a^{c1}_6 +\frac{133}{36} a^{1\ln}_7 +\frac{55}{8} a^{c1}_7 +\frac{949765}{8192}\pi^4\right. \nonumber\\
&& \left. +\frac{20427}{14}\gamma+\frac{238813360021}{3440640}+\frac{41319}{14}\ln(2)\right)\nu^4\nonumber\\
&&+\left(-\frac{557685}{112}\ln(3)+\frac{31955}{48} a^{c1}_6 +\frac{13120}{3}\pi^2\ln(2)-\frac{4}{9} a^{1\ln^2 }_{8} -\frac{205}{8}\pi^2 a^{c1}_6\right.\nonumber\\
&&\left.
 +\frac{84827714750623}{418037760}+\frac{2267813075}{6291456}\pi^4-\frac{28}{9} a^{1\ln}_8 +\frac{43}{6} a^{1\ln}_7 -\frac{1497149729483}{70778880}\pi^2\right.\nonumber\\
&&\left.-\frac{95658095}{1134}\ln(2)+\frac{205}{8} a^{c1}_7 -\frac{50916857}{1134}\gamma-\frac{65}{6} a^{c1}_8 +\frac{6560}{3}\pi^2\gamma\right)\nu^3\nonumber\\
&&+\left(\frac{308243}{28}\pi^2\ln(2)+\frac{36123489}{9856}\ln(3)-\frac{3691195285343}{11266117632000}+\frac{675}{16} a^{c1}_6\right.\nonumber\\
&&\left. -\frac{34723318537}{261954}\gamma+\frac{2862592}{315}\gamma^2-\frac{67823447947}{261954}\ln(2)+a^{1\ln}_8 +5 a^{c 1 }_{9} \right.\nonumber\\
&&\left.+\frac{135}{8} a^{c1}_7 +\frac{2}{3} a^{1\ln}_9 -\frac{49815}{56}\pi^2\ln(3)-\frac{277815950785}{100663296}\pi^4+\frac{9}{4} a^{1\ln}_7\right.\nonumber\\
&&\left. +\frac{11450368}{315}\ln(2)\gamma+\frac{282823}{56}\pi^2\gamma+\frac{15}{2} a^{c1}_8 -\frac{126953125}{114048}\ln(5)\right.\nonumber\\
&&\left.+\frac{11450368}{315}\ln^2(2)+\frac{207691929510749}{2972712960}\pi^2-\frac{13312}{3}\zeta(3)\right)\nu^2\nonumber\\
&&+\left(-\frac{189540}{49}\gamma\ln(3)-\frac{113936420642365}{6442450944}\pi^4+\frac{542238707063245154689}{341738901504000}\right.\nonumber\\
&&\left.+\frac{6087776}{315}\ln^2(2)-\frac{2568359375}{370656}\ln(5)-\frac{3157851462764}{99324225}\ln(2)+\frac{221562973952581}{18496880640}\pi^2\right.\nonumber\\
&&\left.-\frac{1247718915244}{99324225}\gamma+\frac{2555584}{147}\ln(2)\gamma+\frac{1022178637107}{39239200}\ln(3)-\frac{189540}{49}\ln(2)\ln(3)\right.\nonumber\\
&&\left.-\frac{87616}{21}\zeta(3)+\frac{7443104}{2205}\gamma^2-\frac{94770}{49}\ln^2(3)\right)\nu\nonumber\\
e_{\rm 8PN}^{\ln}&=&\frac{155}{6}\nu^5+\left(\frac{55}{8} a^{1\ln}_7 +\frac{83427}{28}\right)\nu^4 \nonumber\\
&& +\left(-\frac{94401221}{2268}+\frac{5494}{3}\pi^2+\frac{205}{8} a^{1\ln}_7 -\frac{56}{9} a^{1\ln^2 }_{8} -\frac{65}{6} a^{1\ln}_8 \right)\nu^3 \nonumber\\
&& +\left(\frac{15}{2} a^{1\ln}_8 +\frac{135}{8} a^{1\ln}_7 +5 a^{1\ln}_9 +\frac{5725184}{315}\ln(2)+\frac{4}{3} a^{1\ln^2 }_{9}\right.\nonumber\\
&&\left. +\frac{282823}{112}\pi^2+2 a^{1\ln^2 }_{8} -\frac{35359866757}{523908}+\frac{2862592}{315}\gamma\right)\nu^2 \nonumber\\
&& +\left(-\frac{94770}{49}\ln(3)+\frac{1277792}{147}\ln(2)+\frac{7443104}{2205}\gamma-\frac{617502707222}{99324225}\right)\nu  \nonumber\\
e_{\rm 8PN}^{\ln^2 }&=&   \frac{15}{2}\nu^2 a^{1\ln^2 }_{8} +\frac{1860776}{2205}\nu+\frac{715648}{315}\nu^2-\frac{65}{6}\nu^3 a^{1\ln^2 }_{8} +5\nu^2 a^{1\ln^2 }_{9}    
\end{eqnarray}
\begin{eqnarray}
e_{\rm 8.5PN}^{c} &=& -\frac{400}{243}\nu^5 a^{c1}_{6.5} +\left(-\frac{219136}{5103}\pi+\frac{80}{9} a^{c1}_{7.5} -\frac{868}{9} a^{c1}_{6.5} \right)\nu^4\nonumber\\
&&+\left(\frac{20812}{27} a^{c1}_{6.5} -\frac{533}{18} a^{c1}_{6.5} \pi^2+\frac{88}{3} a^{c1}_{7.5} -\frac{112}{9} a^{c1}_{8.5} -\frac{124216976}{33075}\pi\right)\nu^3\nonumber\\
&& +\left(\frac{16}{3} a^{c1}_{9.5} -\frac{3649984}{4725}\pi^3+\frac{2}{3} a^{ 1\ln }_{9.5} +8 a^{c1}_{8.5} +\frac{468077708656}{29469825}\pi+45 a^{c1}_{6.5} +18 a^{c1}_{7.5} \right)\nu^2\nonumber\\
&&+\left(\frac{3506176}{4725}\pi^3-\frac{750321664}{165375}\pi\ln(2)+\frac{2734706893326827}{196661965500}\pi-\frac{375160832}{165375}\pi\gamma\right)\nu\nonumber\\
e_{\rm 8.5PN}^{\ln} &=&   -\frac{187580416}{165375}\nu\pi+\frac{16}{3}\nu^2a_{9.5}^{1\ln}  \,.
\end{eqnarray}

\end{widetext}

Let us recall that Ref. \cite{bdtd_beyond} has shown  that, at the $n$PN level (i.e. in $ \nu a_{n+1}(\nu) u^{n+1}$),
the maximum power $p$ of $\nu$ in $A(u;\nu)=1-2u+ \nu a(u;\nu)$ certainly satisfies the inequality
$p \leq n-1$ (which ensures that the values of the coefficients
of $\nu^n$ in $e_{n{\rm PN}}$ explicitly appearing in the formulas above are all exact), and
conjecturally satisfies the stronger inequality  $ 2  p \leq n$.

\noindent {\bf Acknowledgments.}  
D.B. thanks the Italian INFN (Naples) for partial support and IHES for hospitality during the development of this project.
Both  authors are grateful to ICRANet for partial support.

\end{document}